\documentclass[aps,prd,notitlepage,nofootinbib,english,showpacs,twocolumn,10pt]{revtex4-1}

\usepackage[latin1]{inputenc}
\usepackage{float}
\usepackage{latexsym}
\usepackage{graphicx}
\usepackage{amssymb}
\usepackage{amsmath}
\usepackage{amsfonts}
\usepackage{color}
\usepackage{cool}
\usepackage{subfig}
\usepackage{dcolumn}
\usepackage{hyperref}

\begin{document}

\title{Quintessence with Yukawa Interaction}

\author{Andr\'e A. Costa}\email{alencar@if.usp.br}
\author{Lucas C. Olivari}\email{lolivari@if.usp.br}
\author{E. Abdalla}\email{eabdalla@usp.br}
\affiliation{Instituto de F\'isica, Universidade de S\~ao Paulo, C.P.
  66318, 05315-970, S\~ao Paulo, SP, Brazil}
\date{\today}

\begin{abstract}

We consider a quintessence model for dark energy interacting with dark matter via a Yukawa interaction. To put constraints on this model we use the CMB measurements from the Planck satellite together with BAO, SNIa and $H_0$ data. We conclude that this is a viable model and an appropriate scalar potential can favor the interacting scenario.

\end{abstract}

\pacs{98.80.Es, 98.80.Jk, 95.30.Sf}
\maketitle

\section{Introduction}

Several cosmological observations, such as the measurement of the temperature anisotropies in the cosmic microwave background (CMB) \cite{Komatsu:2008hk, Ade:2013ktc, Planck:2013kta, Ade:2013zuv}, the measurement of the apparent magnitude of Type Ia supernovae (SNIa) as a function of redshift \cite{Kowalski:2008ez}, and the measurement of baryon acoustic oscillations (BAO) \cite{Cole:2005sx, Eisenstein:2005su}, have demonstrated that the Universe is currently in an accelerated phase of expansion and that its total energy budget is dominated by a dark energy component. The nature of dark energy is, despite years of intense investigations, an unsolved problem, both under the theoretical and the observational point of view.

The most straightforward candidate for dark energy is the cosmological constant $\Lambda$, which has a constant equation of state parameter $\omega = - 1$. In the standard $\Lambda$-cold dark matter ($\Lambda$CDM) model of the Universe, the cold dark matter only interacts with other components gravitationally, while the dark energy is simply the vacuum energy and therefore has no dynamics. This model fits very well the current observational data, including the recent Planck data \cite{Ade:2013ktc, Planck:2013kta, Ade:2013zuv}. Despite its observational success, this model exhibits some theoretical shortcomings such as the discrepancy between the value of the vacuum energy obtained through observations and the theoretically estimated value \cite{Martin:2012}. This model also suffers from a coincidence problem, i.e., why is the energy densities of matter and dark energy comparable in size at late times \cite{Weinberg:1988cp, Zlatev:1998tr}?

Many alternative models for dark energy that attempt to avoid the problems in the $\Lambda$CDM model have been proposed in the literature. Most of them make use of a dynamical field to describe the dark energy, such as quintessence \cite{Peccei:1987mm, Wetterich:1987fk, Ratra:1987rm} and K-essence \cite{Chiba:1999ka, ArmendarizPicon:2000dh}. Despite the fact that none of these models actually solve the problems that plague the cosmological constant nor provide a better fit to data than $\Lambda$CDM, some strong arguments have been given to justify the use of dynamical dark energy models to describe the Universe \cite{Peccei:1987mm, Wetterich:1987fk, Ratra:1987rm}. 

The quintessence model is composed by a canonical scalar field $\phi$ that slowly rolls down a potential energy $V(\phi)$. In this case, the dark energy has a dynamical equation of state $\omega$ and it can form large scale structures. Also, for being a dynamic component, the quintessence can naturally interact with other components of the Universe, such as the cold dark matter and neutrinos. 

The idea that there is an interaction between dark energy and dark matter has a number of interesting properties from a cosmological point of view. First, it has the theoretically appealing idea that the full dark sector can be treated in a single framework. It can thus help us alleviating the coincidence problem, since the dark energy density now depends on the dark matter energy density. Also an appropriate interaction can accommodate an effective dark energy equation of state in the phantom region at the present time \cite{He:2008si}. At last, the interaction between dark energy and dark matter can affect significantly the expansion history of the Universe and the evolution of density perturbations, which allows us to constrain the parameters of such a model through cosmological observations. 

Cosmologies in which an interaction between dark energy and dark matter is present have been widely explored before in the literature, both at a phenomenological and at a Lagrangian level \cite{Amendola:1999er, Amendola:2003eq, Pavon:2005yx, Boehmer:2008av, Olivares:2005tb, Valiviita:2008iv, Zimdahl:2005bk, Valiviita:2009nu, Xia:2009zzb, Honorez:2010rr, He:2010ta, Bertolami:2007zm, Bertolami:2007tq, Micheletti:2009pk, Micheletti:2009jy, Costa:2013sva}. However, most of the approaches that attempt to discuss an interacting dark sector at a Lagrangian level are built within the framework of modified gravity \cite{Amendola:1999er, Pettorino:2013oxa} or treat the dark energy as an exotic form of matter \cite{Micheletti:2009pk, Micheletti:2009jy}. The model that will be discussed in this work is, on the other hand, built within the framework of the standard quantum field theory in an attempt to be as simple as possible. To accomplish this, we will treat the dark energy as a canonical scalar field, like the scalar field of the quintessence model, and the dark matter as a spin $\frac{1}{2}$ fermionic field. We postulate that the dark energy interacts only with the dark matter. Consequently, in our model, the baryonic matter respects the same conservation equation as it does in the $\Lambda$CDM model. Therefore, locally in the Solar System, where the baryonic matter dominates, we do not have a fifth force problem \cite{Friedman:1991dj,Khoury:2003aq}. Of course, the scalar field will induce a fifth force in the dark matter component which can affect the barionic distribution through gravitation, however this effect will only be important at large distances where dark matter dominates \cite{Friedman:1991dj,Kesden:2006zb,Kesden:2006vz,Honorez:2010rr}. We also postulate that the interaction in the dark sector is given by a Yukawa term which couples the scalar and the fermionic fields. The Yukawa interaction is renormalizable and well studied in the literature, including cosmology \cite{Pavan:2011xn, Farrar:2003uw}. 

In order to constrain the cosmological parameters, we make use of the latest high precision Planck data on CMB temperature anisotropies together with the latest data on BAO, SNIa and the latest constraint on the Hubble constant \cite{Riess:2011yx}. 

This paper is organized as follows: in section~\ref{sec:ym} we describe the interaction model between dark energy and dark matter derived from a Lagragian with a Yukawa coupling and present the background and linear perturbation equations. In section~\ref{sec:analysis} we explain the methods used in the analysis. Section~\ref{sec:results} presents and discusses the results of the analysis. Finally, we summarize our results and conclusions in section~\ref{sec:conclusions}.

\section{The Yukawa Model}
\label{sec:ym}

The coupled dark sector, consisting of a canonical scalar field representing the dark energy and a spin $\frac{1}{2}$ fermionic field representing the dark matter, is described by the Lagrangian
\begin{equation}
\label{lagr}
\mathcal{L} = - \frac{1}{2} \partial^{\mu} \phi \partial_{\mu} \phi - V(\phi) - m(\phi) \bar{\psi} \psi + \mathcal{L}_{\mathrm{K}} [ \psi ]\quad ,
\end{equation}
where $V(\phi)$ is the scalar field potential, which, in principle, can have any functional form, $\mathcal{L}_{\mathrm{K}}$ is the kinetic part of the fermionic Lagrangian, and $m(\phi)$ is the effective fermionic mass, which, in our model, is given by
\begin{equation}
m(\phi) = M - \beta \phi\quad ,
\end{equation}
where $M$ is the fermionic mass and $\beta$ is the Yukawa coupling constant. 

In what follows, we consider that the metric is given by the flat Friedmann-Lemaitre-Robertson-Walker (FLRW) metric, which, when written in terms of the conformal time, $\eta$, is given by
\begin{equation}
\label{frw}
\mathrm{d}s^2 = - a^2(\eta) \mathrm{d} \eta^2 + a^2(\eta) \delta_{ij} \mathrm{d}x^i \mathrm{d} x^j
\end{equation}
and every ``temporal'' derivative is taken with respect to the conformal time.

The conservation equations for the energy densities of dark energy $(d)$ and dark matter $(c)$, which is considered pressureless, are given by
\begin{align}
\label{conserv}
&\dot{\rho}_d = - 3 \mathcal{H} \rho_d (1 + \omega) + Q_0, \nonumber \\ &\dot{\rho}_c = - 3 \mathcal{H} \rho_c - Q_0\quad ,
\end{align}
where $\mathcal{H} = \frac{1}{a} \frac{\mathrm{d} a}{\mathrm{d} \eta}$, $\omega \equiv P/\rho$ is the dark energy equation of state parameter, and $Q_0$ is a generic function representing the exchange of energy in the dark sector. Here we have treated both components of the dark sector as a fluid with the energy-momentum tensor ${T_{A \mu \nu} = (\rho_A + P_A) u_{A \mu} u_{A \nu} + P_A g_{\mu \nu}}$, where $u_{A \mu} = (- a, 0, 0, 0)$ is the $A$-fluid 4-velocity. From the Lagrangian, Eq. $\eqref{lagr}$, the scalar field has energy density and pressure given by
\begin{equation}
\label{scalar}
\rho_d = \frac{\dot{\phi}^2}{2 a^2} + V(\phi), \; \; \; P_d = \frac{\dot{\phi}^2}{2 a^2} - V(\phi)\quad .
\end{equation}

The source $Q_0$ that appears in the energy conservation equations, Eqs. $\eqref{conserv}$, is related to the effective fermionic mass appearing in the Lagrangian by the relation \cite{Amendola:2003wa, Pettorino:2008ez}
\begin{equation}
Q_{\mu} = - \frac{\partial \ln m(\phi)}{\partial \phi} \rho_c \nabla_{\mu} \phi\quad .
\end{equation} To obtain this relation it is necessary to use the equations of motion of the scalar field and the fermionic field, which can be obtained from the Lagrangian through the variational principle, and the supposition that these fields can be described by perfect fluids in a cosmological level \cite{Amendola:2003wa, Pettorino:2008ez}. Therefore the model that we consider here corresponds to the choice
\begin{align}
\label{Q_lagrangian}
Q_0 &= \frac{\beta}{M - \beta \phi} \rho_c \dot{\phi} \nonumber \\ &= \frac{r}{1 - r \phi} \rho_c \dot{\phi}\quad ,
\end{align}
where we have defined $r \equiv \frac{\beta}{M}M_{pl}$ and we chose to normalize all mass scales with the reduced Planck mass $M_{pl} = 1/\sqrt{8 \pi G} = 2.435 \times 10^{18}$ GeV, where $G$ is the gravitational constant. We see from equation $\eqref{Q_lagrangian}$ that the interaction $\beta$ and the fermion mass $M$ are degenerate and so we cannot know both at the same time but only the ratio $r$. Therefore, we use $r$ instead of $\beta$ as our interaction parameter. This has the advantage of decreasing one degree of freedom in the analysis, at the cost that we are unable to know the individual values of $\beta$ or $M$. 

We can generalize the energy-momentum conservation equations for the dark sector components to the covariant form
\begin{align}
\label{conser}
&\nabla_{\nu} T_{d \mu}^{\nu} = \frac{r}{1 - r \phi} \rho_c \nabla_{\mu} \phi, \nonumber \\ &\nabla_{\nu} T_{c \mu}^{\nu} = - \frac{r}{1 - r \phi} \rho_c \nabla_{\mu} \phi\quad .
\end{align}

\subsection{Background Evolution}

As we have previously assumed that the background Universe is described by a flat FLRW metric, Eq. $\eqref{frw}$, we are led, by the Einstein field equations, to the Friedmann equation,
\begin{equation}
\label{fried}
\mathcal{H}^2 \equiv \left(\frac{\dot{a}}{a} \right)^2 = \frac{8 \pi G}{3} a^2 \rho_t\quad , 
\end{equation}
where $\rho_t$ is the total energy density. Using equation $\eqref{scalar}$, the Friedmann equation $\eqref{fried}$ can be written as
\begin{equation}
\mathcal{H}^2 = \frac{8 \pi G}{3} a^2 \left( \rho_r + \rho_b + \rho_c + \frac{\dot{\phi}^2}{2 a^2} + V(\phi) \right)\quad .
\end{equation}
In this equation we are considering all the components of the Universe, with $\rho_r$ and $\rho_b$ being the radiation (photons and neutrinos) and baryonic energy densities, respectively. We can define the energy density parameters $\Omega_A \equiv \rho_A/\rho_t$, where $\rho_A$ is the energy density of the $A$-fluid.

Considering that the dark sector of the Universe respects the characteristic equations of our model, Eqs. $\eqref{conser}$, for all the components of the Universe, the time components of the energy-momentum conservation equation are
\begin{align}
&\dot{\rho}_r + 4 \mathcal{H} \rho_r = 0, \\ &\dot{\rho}_b + 3 \mathcal{H} \rho_b = 0, \\ &\dot{\rho}_c + 3 \mathcal{H} \rho_c = - \frac{r}{1 - r \phi} \rho_c \dot{\phi}, \\ \label{deeq} &\dot{\rho}_d + 3 \mathcal{H} \rho_d (1 + \omega) = \frac{r}{1 - r \phi} \rho_c \dot{\phi}\quad .
\end{align}
Since the dark energy equation of state, $\omega$, and the interaction, $Q_0$, depend on the scalar field and its conformal time derivative, $\phi$ and $\dot{\phi}$, we use the Klein-Gordon equation, which is obtained directly from the Lagrangian, Eq. $\eqref{lagr}$, to completely describe the dark energy component,
\begin{equation}
\ddot{\phi} + 2 \mathcal{H} \dot{\phi} + a^2 V'(\phi) = a^2 \frac{r}{1 - r \phi} \rho_c\quad  ,
\end{equation}
where the prime denotes a derivative with respect to the scalar field $\phi$.

The solution of these equations is obtained choosing initial conditions such that the energy density parameters today are given by the observed parameters $\Omega_c$ and $\Omega_d$. Thus, we have three initial conditions, $\rho_{c_i}$, $\phi_i$ and $\dot{\phi}_i$, to satisfy two parameters. Without loss of generality, we can fix $\dot{\phi}_i$. In fact, we have varied $\dot{\phi}_i$ over a large range and verified that the results are independent of it.

The interaction $Q_0$ has a diverging point at $r\phi = 1$, thus to avoid this point we will stay in the region $r\phi < 1$. Therefore, the sign of the interaction depends on $r$ and $\dot{\phi}$, if they have the same sign, $Q_0$ is positive and we have a flux of energy from dark matter to dark energy. On the other hand, if $r$ and $\dot{\phi}$ have opposite signs, $Q_0$ is negative and the flux is from dark energy to dark matter. We observe that, unlike the phenomenological case \cite{Costa:2013sva}, the sign of $Q_0$ is not necessary fixed, it can change during the evolution of the Universe.

\subsection{Linear Perturbations}

In this section we will consider the evolution of linear cosmological perturbations in our model. In the synchronous gauge, the line element of the linearly perturbed FLRW metric is given by \cite{Kodama:1985bj, Mukhanov:1990me}
\begin{equation}
\label{sinc}
\mathrm{d} s^2 = - a^2(\eta) \mathrm{d} \eta^2 + a^2(\eta)[(1 + \frac{1}{3} h) \delta_{ij} + D_{ij} \chi] \mathrm{d}x^i \mathrm{d}x^j\quad .
\end{equation}
Here, we will restrict our analysis to the scalar modes, $h$ and $\chi$, of the metric perturbations.

The inhomogeneous part of the energy density of dark matter and the scalar field can be written as
\begin{align}
&\rho_c(\eta, \vec{x}) = \rho_c(\eta) [1 + \delta_c(\eta, \vec{x})], \\ &\phi(\eta, \vec{x}) = \phi(\eta) + \varphi(\eta, \vec{x})\quad ,
\end{align}
where $\rho_c(\eta)$ and $\phi(\eta)$ concern the background while $\delta_c$ and $\varphi$ are the linear perturbations. For the dark matter, using the perturbed part of the energy-momentum conservation equations for the dark sector, Eqs. $\eqref{conser}$, we obtain, in the Fourier space, the following equations,
\begin{align}
&\dot{\delta}_c = - \theta_c - \frac{\dot{h}}{2} - \frac{r}{1 - r \phi} \dot{\varphi} + \frac{r^2}{(1 - r \phi)^2} \dot{\phi} \varphi, \\  &\dot{\theta_c} = - \mathcal{H} \theta_c + \frac{r}{1 - r \phi} \theta_c \dot{\phi} - k^2 \frac{r}{1 - r \phi} \varphi\quad ,
\end{align}
where $\theta_c = i k_j v_c^j$ is the gradient of the velocity field. In these equations, we have neglected the shear stress of the dark matter, which is always small because of its non-relativistic character. We note that, in the presence of the interaction, the gradient of the velocity $\theta_c$ will be non-zero throughout the Universe evolution. This means that, instead of working in the cold dark matter rest frame, we will work in an arbitrary synchronous gauge.

For the dark energy, we only need the time component of the energy-momentum conservation equation $\eqref{conser}$, which gives
\begin{widetext}
\begin{equation}
\ddot{\varphi} + 2 \mathcal{H} \dot{\varphi} + k^2 \varphi + a^2 \frac{\mathrm{d}^2 V}{\mathrm{d} \phi^2} \varphi + \frac{\dot{h} \dot{\phi}}{2} = - a^2 \frac{r^2}{(1 - r \phi)^2} \varphi \rho_c + a^2 \frac{r}{1 - r \phi}\rho_c \delta_c\quad .
\end{equation}
\end{widetext}
To the other components of the Universe, baryons and radiation, we have the same perturbed equations as in the $\Lambda$CDM model.

\begin{figure*}[]
\centering \subfloat[
]{\includegraphics[width=0.35\textwidth,angle=-90]{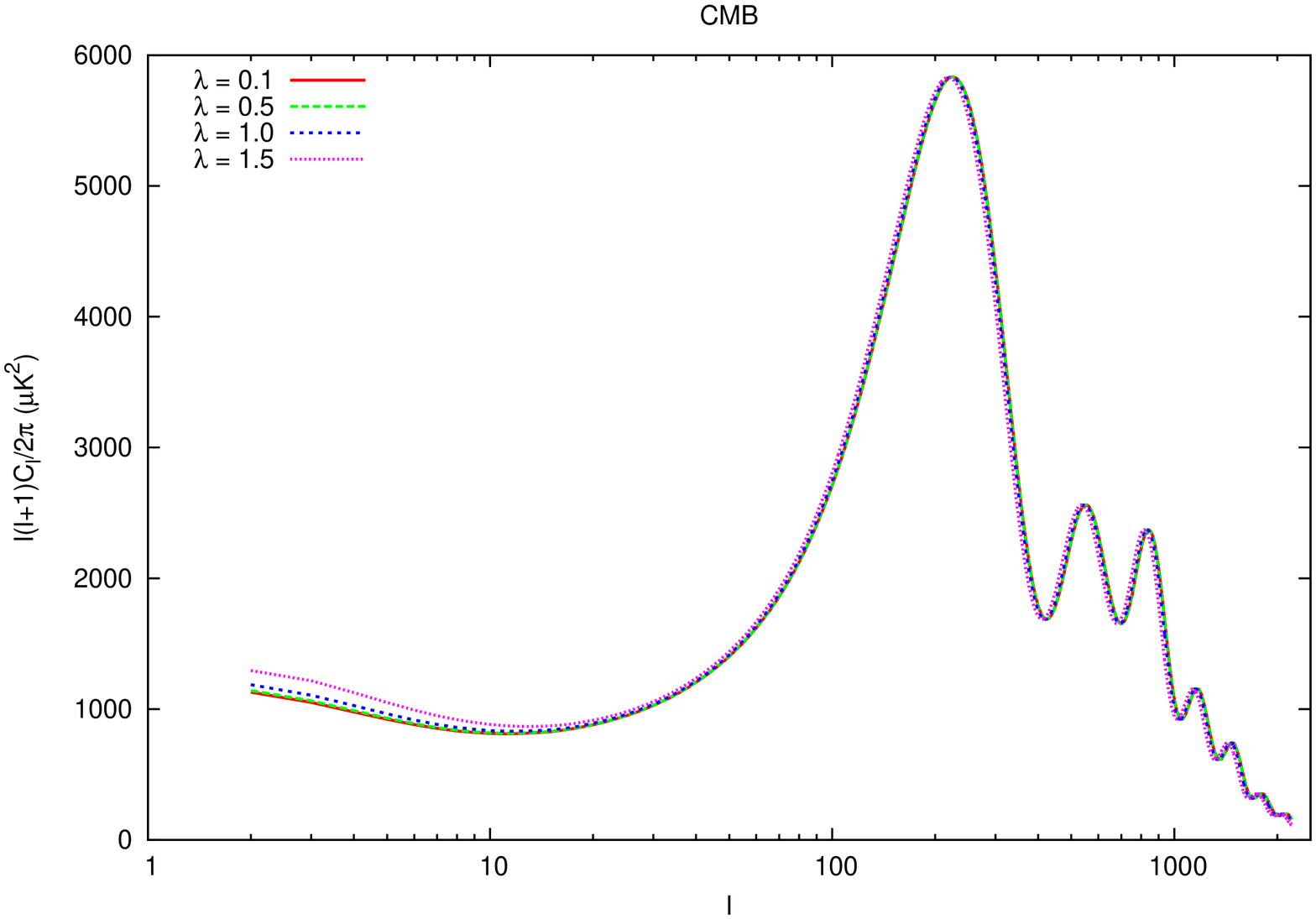}}
\subfloat[
]{\includegraphics[width=0.35\textwidth,angle=-90]{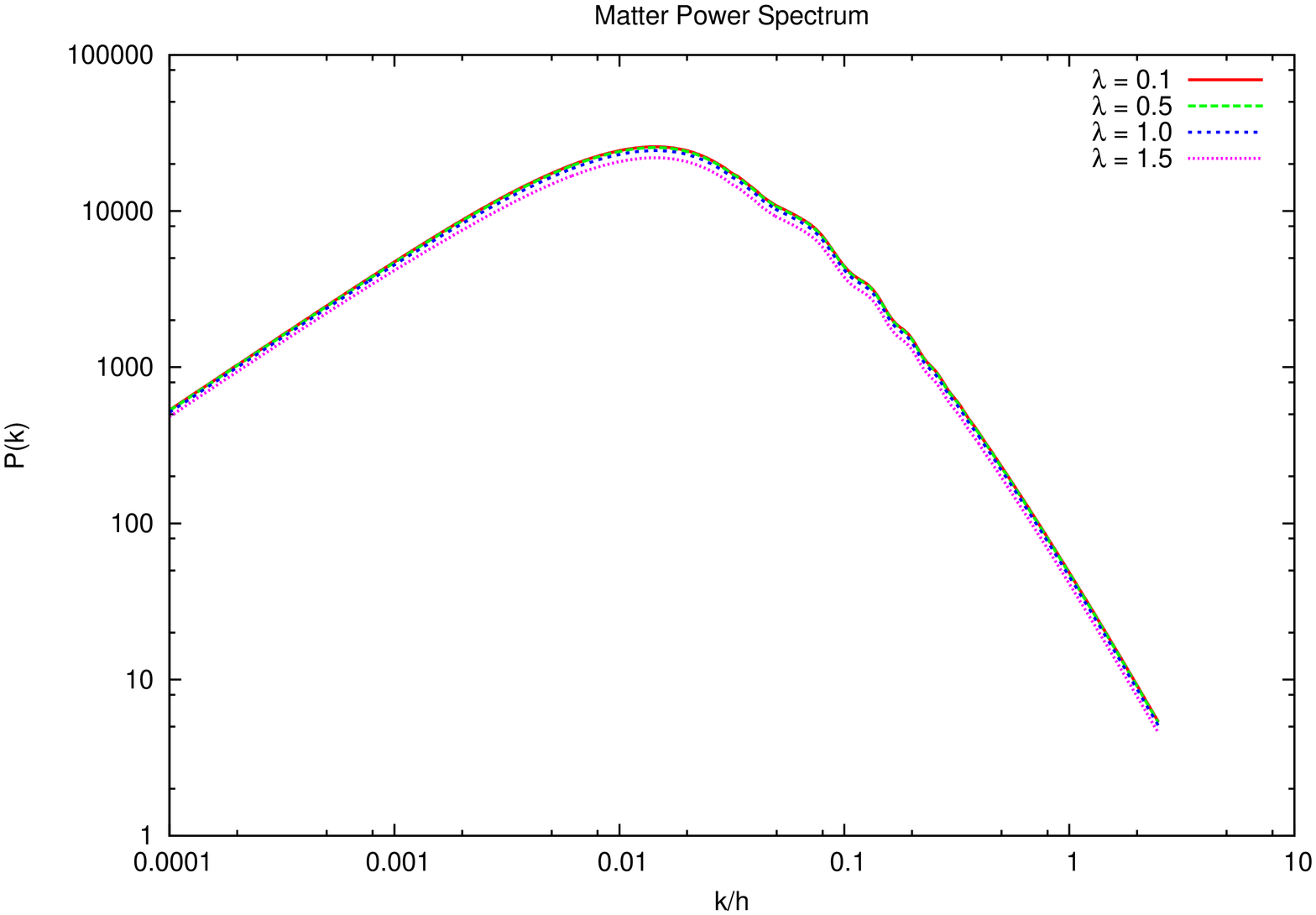}}
\caption{(Color online). Power spectra for the Yukawa model with $r = \frac{\beta}{M}M_{pl} = 0$ and different values of the scalar potential parameter.} \label{Spectra_Qcouple1}
\end{figure*}

\begin{figure*}[]
\centering \subfloat[
]{\includegraphics[width=0.35\textwidth,angle=-90]{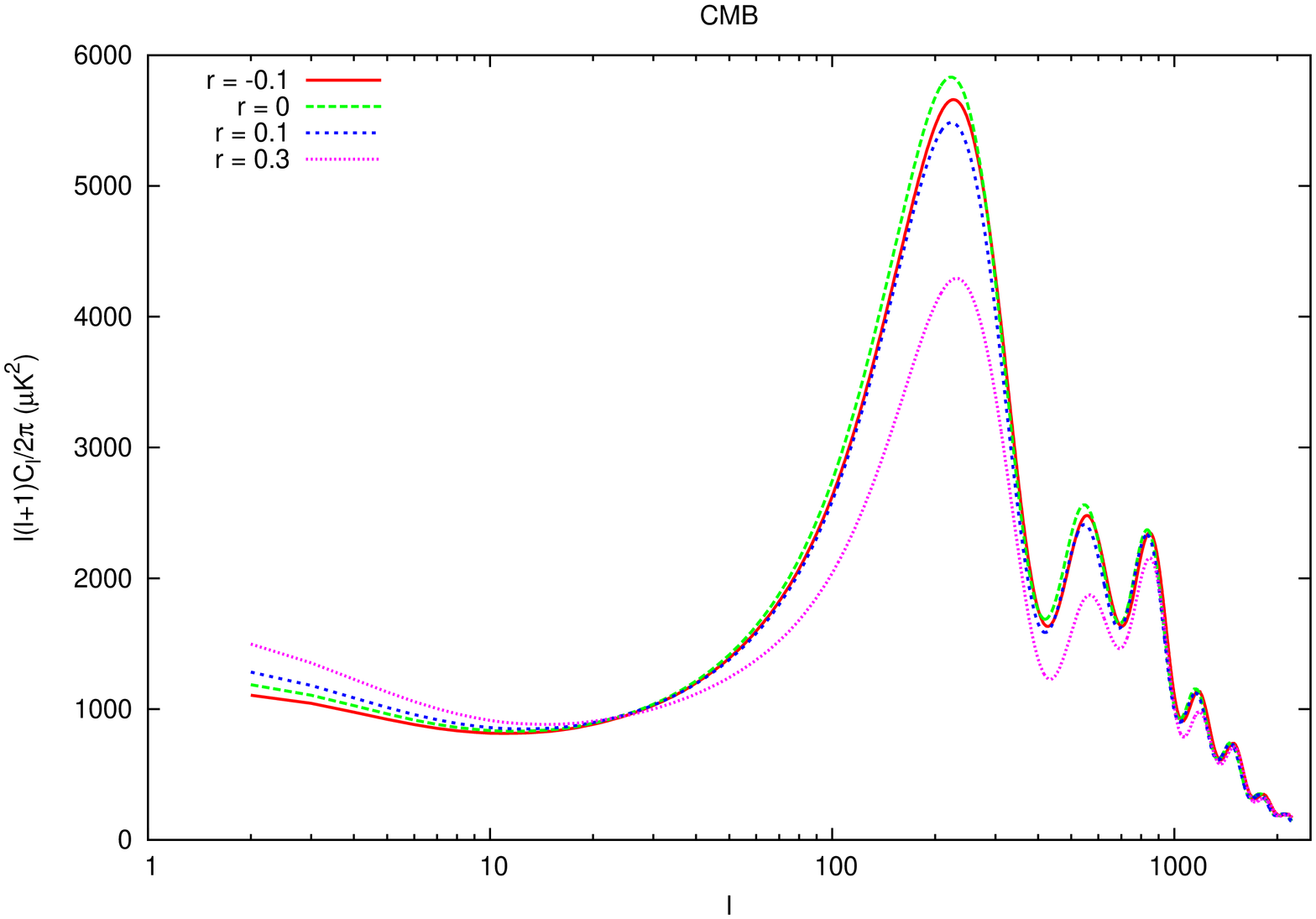}}
\subfloat[
]{\includegraphics[width=0.35\textwidth,angle=-90]{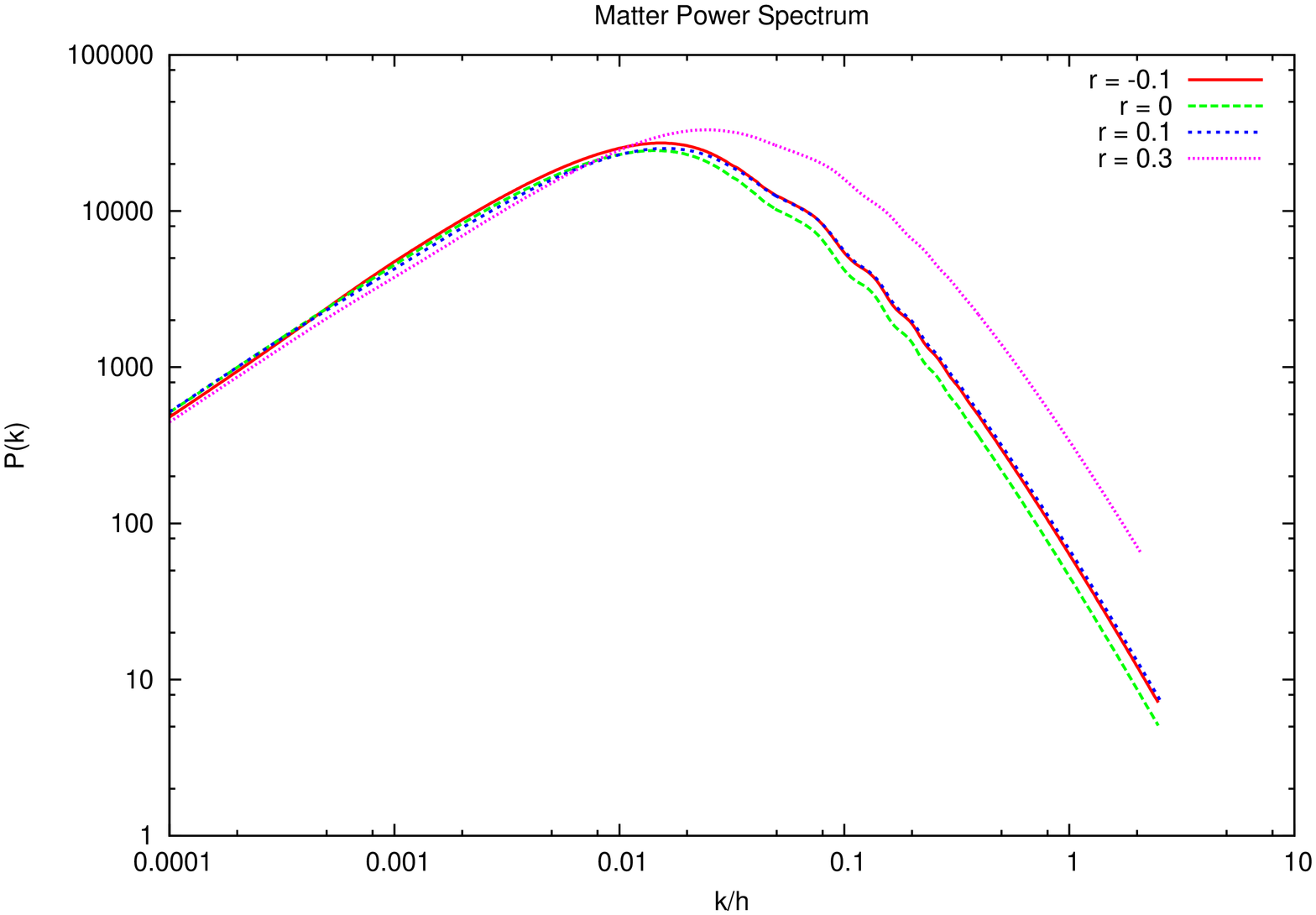}}
\caption{(Color online). Power spectra for the Yukawa model with $\lambda = 1$ and different values of the dimensioless interaction parameter $r = \frac{\beta}{M}M_{pl}$.} \label{Spectra_Qcouple2}
\end{figure*}

To solve these perturbed equations we need to provide two initial conditions to $\varphi$ and $\dot{\varphi}$. We will consider that, at early times, we have adiabatic initial conditions for the dark sector, which implies that \cite{Valiviita:2008iv}
\begin{equation}
\frac{\delta \rho_{\phi}}{\rho_{\phi} + P_{\phi}} = \frac{\delta \rho_c}{\rho_c + P_c}\quad ,
\end{equation}
and that the scalar field intrinsic perturbation is zero,
\begin{equation}
\frac{\delta \rho_{\phi}}{\rho_{\phi}} - \frac{\delta P_{\phi}}{P_{\phi}} = 0\quad .
\end{equation}
These choices are not determinant to the Universe evolution because isocurvature perturbations will be produced due to the presence of non-minimal coupling in the dark sector \cite{Chongchitnan:2008ry}.

We modify the CAMB code \cite{Lewis:1999bs} to include the Lagrangian model above. We consider that the scalar potential is given by
\begin{equation}
V(\phi) = Ae^{-\lambda\phi/M_{pl}} \quad ,
\end{equation}
where $A$ is a normalization constant and $\lambda$ is a dimensionless parameter. We set $A$ to be equal to the value of the cosmological constant energy density $A = \rho_\Lambda$ in the $\Lambda$CDM cosmology. Thus, $\lambda \neq 0$ and $r \neq 0$ is a measure of how our model differs from the cosmological constant model. 

Figures \ref{Spectra_Qcouple1} and \ref{Spectra_Qcouple2} present some graphs for the CMB and matter power spectrum obtained from our interacting model for different values of its parameters. Figure \ref{Spectra_Qcouple1} shows that the scalar potential parameter $\lambda$ has a small effect on the CMB and matter power spectrum, affecting mainly the low-$l$ CMB power spectrum. On the other hand, we see from Fig. \ref{Spectra_Qcouple2} that, in addition to modifying the CMB spectrum at low $l$, the coupling between dark matter and dark energy influences the acoustic peaks at large multipoles.

\section{Data Analysis}
\label{sec:analysis}
To constrain the cosmological parameters in our interacting model, we use several data sets: the measurements of CMB anisotropies, BAO, SNIa and the direct measurement of the Hubble constant $H_0$. Below we describe the likelihood for these measurements.

The Planck data set that we use is a combination of
the low-$l$ TT likelihood, which includes measurements for $l
= 2-49$, combined with the high-$l$ TT likelihood, which includes
measurements from $l = 50$ up to a maximum multipole number of
$l_{\mathrm{max}} = 2500$ \cite{Ade:2013ktc,Ade:2013zuv,Planck:2013kta}.
Together with the Planck data, we include the
polarization measurements from the nine year
Wilkinson Microwave Anisotropy Probe (WMAP)
\cite{Bennett:2012zja}, the low-$l$ ($l<32$) TE,
EE, and BB likelihoods.

In addition to the CMB data sets, we also
consider measurements of baryon acoustic oscillations (BAO)
in the matter power spectrum. We combine the results from three
redshift surveys: the 6dF Galaxy Survey measurement at redshift $z = 0.106$ \cite{Beutler:2011hx}, the SDSS DR7 BAO measurement at redshift $z = 0.35$, as analysed by Padmanabhan et al.
\cite{Padmanabhan:2012hf} and the BOSS DR9 measurement at $z =
0.57$ \cite{Anderson:2012sa}. These redshift surveys measure the distance ratio
\begin{equation}
d_z = \frac{r_s(z_{\mathrm{drag}})}{D_V(z)} \quad ,
\end{equation}
where $r_s(z_{\mathrm{drag}})$ is the comoving sound horizon at the baryon drag epoch, which is the epoch when baryons became dynamically decoupled from photons, and $D_V(z)$ combines the angular diameter distance $d_A(z)$ and the Hubble parameter $H(z)$, in a way appropriate for the analysis of spherically-averaged two-point statistics,
\begin{equation}
D_V(z) = \left[(1 + z)^2d_A^2(z)\frac{cz}{H(z)}\right]^{1/3}\quad .
\end{equation}
The comparison with BAO measurements is made using $\chi^2$ statistics
\begin{equation}
\chi^2_{\mathrm{BAO}} = ({\bf x} - {\bf x}^{\mathrm{obs}})^T C^{-1}_{\mathrm{BAO}}({\bf x} - {\bf x}^{\mathrm{obs}}) \quad ,
\end{equation}
where ${\bf x}$ is our theoretical predictions and ${\bf x}^{\mathrm{obs}}$ denotes the data vector. The data vector is composed of the measurements of the three data sets above: for the 6dF, $D_V(0.106) = (457 \pm 27)$ Mpc, for the DR7, $D_V(0.35)/r_s = 8.88 \pm 0.17$, and for the DR9, $D_V(0.57)/r_s = 13.67 \pm 0.22$.

We also use the SNIa data from the Supernova Cosmology Project (SCP) Union 2.1 compilation \cite{Suzuki:2011hu}, which has 580
samples. The Union 2.1 uses SALT2 \cite{Guy:2007dv} to fit supernova lightcurves. The SALT2 model fits three parameters to each supernova: an overall normalization, $x_0$, to the time dependent spectral energy distribution of a SNIa, the deviation from the average lightcurve shape, $x_1$, and the deviation from the mean SNIa B - V color, $c$. Combining these parameters, the distance modulus is given by
\begin{equation}
\mu_B = m_B^{\mathrm{max}} + \alpha\cdot x_1- \beta\cdot c + \delta\cdot P(m_\star^{\mathrm{true}} < m_\star^{\mathrm{threshold}}) - M_B ,
\end{equation}
where $m_B^{\mathrm{max}}$ is the integrated B-band flux at maximum light, $P(m_\star^{\mathrm{true}} < m_\star^{\mathrm{threshold}})$ gives the correlation of SNIa luminosity to the mass of the host galaxy, and $M_B$ is the absolute B-band magnitude. The nuisance parameters $\alpha$, $\beta$, $\delta$ and $M_B$ are fitted simultaneously with cosmological parameters. The best-fit cosmology is determined by minimizing the $\chi^2$,
\begin{equation}
\chi_{SN}^2 = \sum_{i = 1}^{580}\frac{[\mu_B(\alpha, \beta, \delta, M_B) - \mu(z, \Omega_m, \Omega_d, \lambda, r)]^2}{\sigma^2} .
\end{equation}
To test our interacting dark energy model we use the CosmoMC \cite{Lewis:2002ah,Lewis:2013hha} module associated with the Union 2.1 sample. In this module the nuisance parameters are hold fixed with values $\alpha = 0.1218$, $\beta = 2.4657$ and $\delta = -0.03634$.

From observations of Cepheid variables and low-redshift Type Ia surpernovae, the Hubble Space Telescope (HST) determined the Hubble constant with 3.3\% uncertainty, including systematic errors \cite{Riess:2011yx},
\begin{equation}
H_0 = 73.8 \pm 2.4 \; \mathrm{km} \mathrm{s}^{-1} \mathrm{Mpc}^{-1} \quad .
\end{equation}
We use this measurement of the Hubble constant as an additional constraint.

\section{Results}
\label{sec:results}

\begin{table}[htb!]
\centering \caption{Priors for the cosmological
parameters considered in the analysis of
the Yukawa interacting model.}
\begin{tabular}{|c|c|}
\toprule
Parameters & Prior \\
\hline
$\Omega_b h^2$ & $[0.005, 0.1] $\\
\hline
$\Omega_c h^2$ & $[0.001, 0.99]$ \\
\hline
$100 \theta$ & $[0.5, 10]$ \\
\hline
$\tau$ & $[0.01, 0.8]$ \\
\hline
$n_s$ & $[0.9, 1.1]$ \\
\hline
$\ln(10^{10} A_s)$ & $[2.7, 4]$ \\
\hline
$\lambda$ & $[0.1, 1.5]$ \\
\hline
$r = \frac{\beta}{M}M_{pl}$ & $[-0.1, 0.1]$ \\
\lasthline
\end{tabular}
\label{parameters_lagrangian}
\end{table}

\begin{table*}[ht!]
\centering
\caption{Cosmological parameters - Yukawa model}
\label{bestfit_lagrangian}
\resizebox{\textwidth}{!}{
\begin{tabular}{ccccccc}
   \toprule
    & \multicolumn{2}{c}{Planck} & \multicolumn{2}{c}{Planck+BAO} & \multicolumn{2}{c}{Planck+BAO+SNIa+H0} \\
    \cline{2-7}
    Parameter & Best fit  & 68\% limits & Best fit  & 68\% limits & Best fit  & 68\% limits \\
    \hline
$\Omega_b h^2$ & 0.02186 & $0.02195^{+0.000279}_{-0.00028}$ & 0.02208 & $0.02198^{+0.000268}_{-0.000266}$ & 0.02208 & $0.022^{+0.00027}_{-0.000276}$\\
$\Omega_c h^2$ & 0.1159 & $0.1171^{+0.00477}_{-0.00315}$ & 0.1169 & $0.1175^{+0.00252}_{-0.0018}$ & 0.117 & $0.1162^{+0.00218}_{-0.00174}$\\
$100\theta_{MC}$ & 1.041 & $1.041^{+0.000651}_{-0.000645}$ & 1.042 & $1.041^{+0.000578}_{-0.000584}$ & 1.041 & $1.041^{+0.000572}_{-0.000581}$\\
$\tau$ & 0.08589 & $0.08879^{+0.0125}_{-0.0139}$ & 0.08217 & $0.08979^{+0.0121}_{-0.0135}$ & 0.09744 & $0.08998^{+0.0122}_{-0.0137}$\\
$n_s$ & 0.9589 & $0.959^{+0.0075}_{-0.00753}$ & 0.9617 & $0.9594^{+0.00604}_{-0.00603}$ & 0.9626 & $0.9603^{+0.00603}_{-0.00599}$\\
${\rm{ln}}(10^{10}A_s)$ & 3.084 & $3.086^{+0.0246}_{-0.0249}$ & 3.069 & $3.087^{+0.0236}_{-0.026}$ & 3.1 & $3.087^{+0.0245}_{-0.0244}$\\
$\lambda$ & 0.5627 & $0.7497^{+0.75}_{-0.65}$ & 0.1799 & $0.6518^{+0.173}_{-0.552}$ & 0.2435 & $0.4043^{+0.0765}_{-0.304}$\\
$r$ & -0.06695 & $-0.009795^{+0.046}_{-0.0613}$ & 0.02099 & $-0.006231^{+0.0463}_{-0.0548}$ & -0.03865 & $-0.01021^{+0.0794}_{-0.0675}$\\
\hline
$\Omega_d$ & 0.7175 & $0.6882^{+0.03}_{-0.037}$ & 0.7005 & $0.6887^{+0.0167}_{-0.0139}$ & 0.7049 & $0.7069^{+0.012}_{-0.012}$\\
$\Omega_m$ & 0.2825 & $0.3118^{+0.037}_{-0.03}$ & 0.2995 & $0.3113^{+0.0139}_{-0.0167}$ & 0.2951 & $0.2931^{+0.012}_{-0.012}$\\
$z_{re}$ & 10.71 & $10.94^{+1.08}_{-1.08}$ & 10.33 & $11.02^{+1.06}_{-1.05}$ & 11.65 & $11.01^{+1.08}_{-1.08}$\\
$H_0$ & 69.99 & $67.16^{+2.41}_{-3.3}$ & 68.28 & $67.13^{+1.61}_{-1.21}$ & 68.8 & $68.85^{+1.06}_{-1.06}$\\
${\rm{Age}}/{\rm{Gyr}}$ & 13.65 & $13.75^{+0.157}_{-0.0744}$ & 13.78 & $13.77^{+0.071}_{-0.0502}$ & 13.74 & $13.72^{+0.0753}_{-0.0604}$\\
\hline
$\chi^2_{min}/2$ & \multicolumn{2}{c}{4903.10} & \multicolumn{2}{c}{4904.04} & \multicolumn{2}{c}{4971.32} \\
   \lasthline
\end{tabular}
}
\end{table*}

\begin{table*}[ht!]
\centering
\caption{Cosmological parameters - Yukawa model with fixed $\lambda = \sqrt{3/2}$}
\label{bestfit_lagrangian2}
\resizebox{\textwidth}{!}{
\begin{tabular}{ccccccc}
   \toprule
    & \multicolumn{2}{c}{Planck} & \multicolumn{2}{c}{Planck+BAO} & \multicolumn{2}{c}{Planck+BAO+SNIa+H0} \\
    \cline{2-7}
    Parameter & Best fit  & 68\% limits & Best fit  & 68\% limits & Best fit  & 68\% limits \\
    \hline
$\Omega_b h^2$ & 0.02197 & $0.02194^{+0.000273}_{-0.000281}$ & 0.022 & $0.022^{+0.00028}_{-0.000286}$ & 0.02206 & $0.02194^{+0.000284}_{-0.000281}$\\
$\Omega_c h^2$ & 0.1166 & $0.1176^{+0.00421}_{-0.00302}$ & 0.1187 & $0.117^{+0.00343}_{-0.00176}$ & 0.1174 & $0.117^{+0.00166}_{-0.00164}$\\
$100\theta_{MC}$ & 1.041 & $1.041^{+0.000638}_{-0.000636}$ & 1.041 & $1.041^{+0.000612}_{-0.000601}$ & 1.041 & $1.041^{+0.000606}_{-0.000598}$\\
$\tau$ & 0.0931 & $0.08825^{+0.012}_{-0.014}$ & 0.08808 & $0.08996^{+0.0125}_{-0.0139}$ & 0.08497 & $0.08854^{+0.0122}_{-0.0142}$\\
$n_s$ & 0.966 & $0.959^{+0.00734}_{-0.00737}$ & 0.9587 & $0.9607^{+0.0063}_{-0.00632}$ & 0.96 & $0.9598^{+0.0064}_{-0.00644}$\\
${\rm{ln}}(10^{10}A_s)$ & 3.093 & $3.085^{+0.0234}_{-0.0261}$ & 3.086 & $3.086^{+0.0247}_{-0.0252}$ & 3.08 & $3.086^{+0.0237}_{-0.0272}$\\
$r$ & -0.06097 & $-0.01971^{+0.0411}_{-0.0572}$ & -0.04986 & $-0.02005^{+0.021}_{-0.058}$ & -0.07076 & $-0.0669^{+0.0121}_{-0.0175}$\\
\hline
$\Omega_d$ & 0.6887 & $0.6681^{+0.0208}_{-0.0343}$ & 0.673 & $0.6719^{+0.013}_{-0.0129}$ & 0.6919 & $0.691^{+0.0122}_{-0.011}$\\
$\Omega_m$ & 0.3113 & $0.3319^{+0.0343}_{-0.0208}$ & 0.327 & $0.3281^{+0.0129}_{-0.013}$ & 0.3081 & $0.309^{+0.011}_{-0.0122}$\\
$z_{re}$ & 11.28 & $10.9^{+1.07}_{-1.07}$ & 10.89 & $11.02^{+1.09}_{-1.08}$ & 10.57 & $10.88^{+1.1}_{-1.1}$\\
$H_0$ & 66.88 & $65.13^{+1.11}_{-2.76}$ & 65.74 & $65.28^{+0.93}_{-1.28}$ & 67.44 & $67.24^{+0.978}_{-0.997}$\\
${\rm{Age}}/{\rm{Gyr}}$ & 13.7 & $13.78^{+0.159}_{-0.0636}$ & 13.75 & $13.78^{+0.0819}_{-0.0623}$ & 13.63 & $13.66^{+0.0705}_{-0.0708}$\\
\hline
$\chi^2_{min}/2$ & \multicolumn{2}{c}{4903.02} & \multicolumn{2}{c}{4903.90} & \multicolumn{2}{c}{4974.54} \\
   \lasthline
\end{tabular}
}
\end{table*}

We want to put constraints on the cosmological parameters and verify if the Yukawa interaction is favored by the observational data. The priors that we use are listed in Table \ref{parameters_lagrangian}. We observe that the prior we used for $r$ is consistent with the bounds found in the literature \cite{Friedman:1991dj,Kesden:2006zb,Kesden:2006vz,Honorez:2010rr}. At first we allow the parameter of the scalar potential $\lambda$ to vary freely. We fixed the helium abundance at $Y_p = 0.24$. The number of relativistic degrees of freedom is adjusted to $N_{eff} = 3.046$ and the total neutrino mass is set to $\sum m_\nu = 0.06$ eV. At last, the spectrum lensing normalization is $A_L = 1$. To finish the MCMC we set the Gelman and Rubin criterion to $R - 1 = 0.03$ \cite{Gelman:1992}.

\begin{figure*}[htbp!]
\centering
    \includegraphics[scale=0.45]{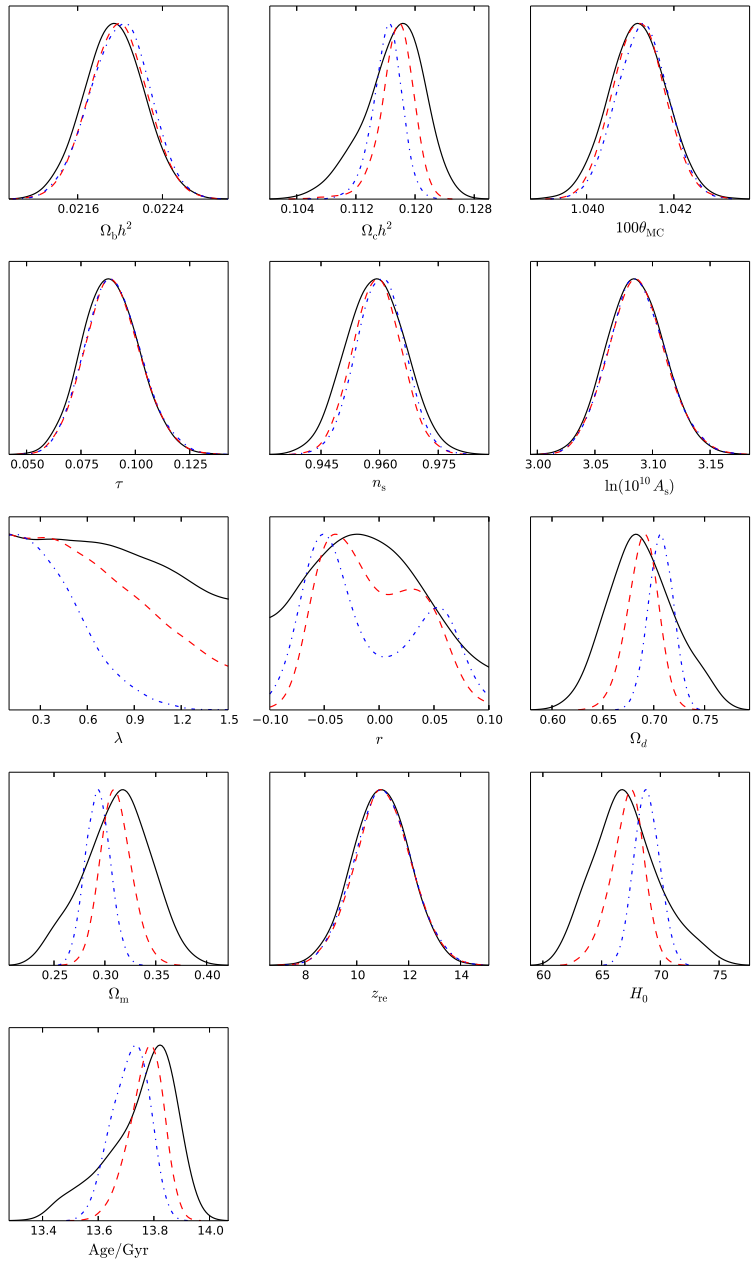}
\caption{(Color online). The posterior for the parameters of the Yukawa model. The black solid lines correspond to the Planck constraints, the red dashed lines correspond to Planck + BAO and the blue dot-dashed lines correspond to Planck + BAO + SNIa + $H_0$.}\label{1d_dist_lagrangian}
\end{figure*}

\begin{figure*}[htbp!]
\centering
    \includegraphics[width=\textwidth]{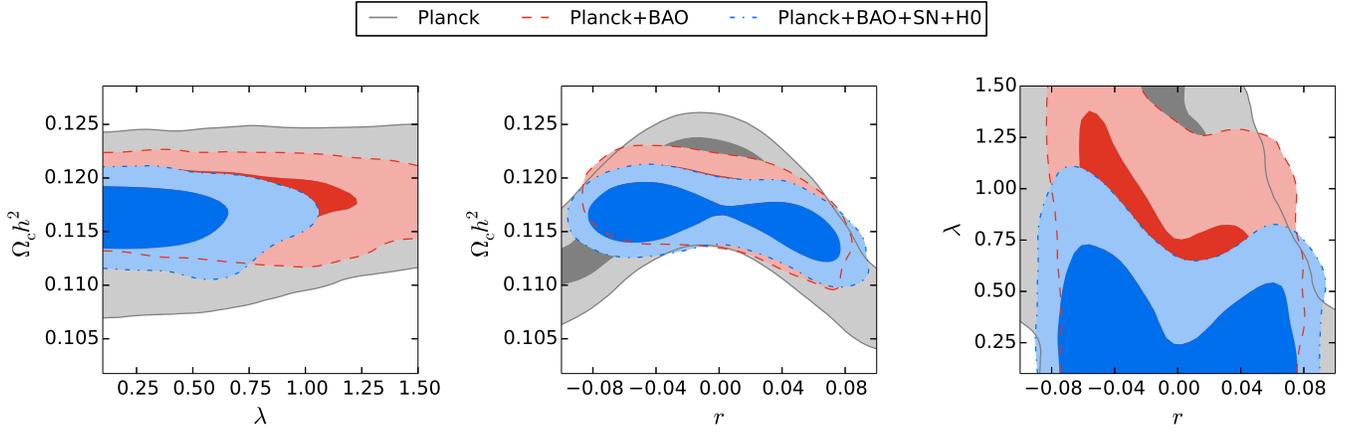}
\caption{(Color online). 2-D distribution for selected parameters - Yukawa model.}\label{2d_dist_lagrangian}
\end{figure*}

\begin{figure*}[htbp!]
\centering
    \includegraphics[scale=0.45]{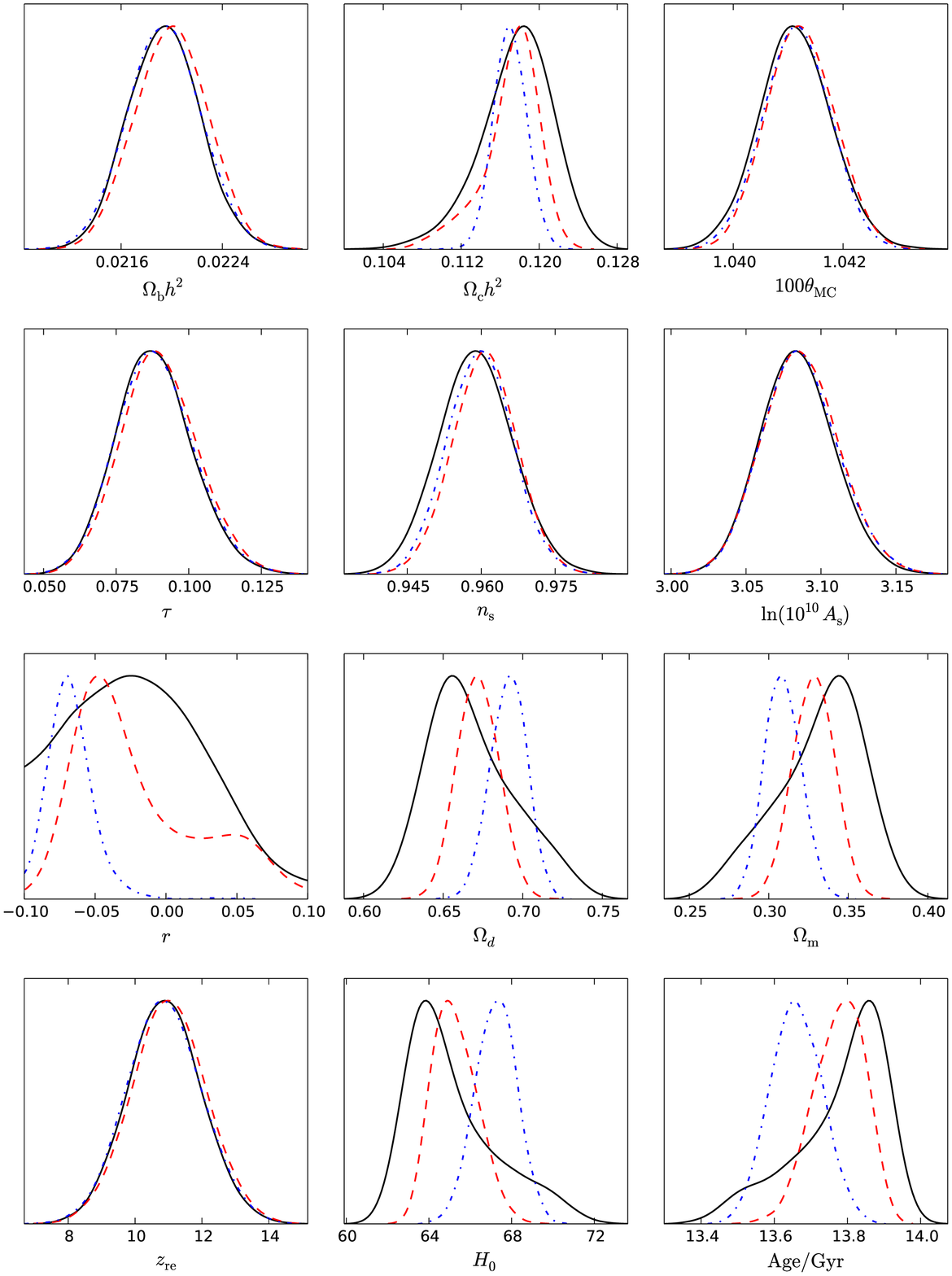}
\caption{(Color online). The posterior for the parameters of the Yukawa model with fixed $\lambda = \sqrt{3/2}$. The black solid lines correspond to the Planck constraints, the red dashed lines correspond to Planck + BAO and the blue dot-dashed lines correspond to Planck + BAO + SNIa + $H_0$.}\label{1d_dist_lagrangian2}
\end{figure*}

We use the measurements of the CMB anisotropies made by Planck together with BAO, SNIa and $H_0$ measurements. Using the priors listed in Table \ref{parameters_lagrangian} we run the MCMC. The results are shown in Table \ref{bestfit_lagrangian}, the 1-D posteriors for the parameters are given in Fig. \ref{1d_dist_lagrangian}, and some parameter degeneracies are given in Fig. \ref{2d_dist_lagrangian}. We observe that the Planck data set alone is not enough to constrain the scalar potential parameter $\lambda$ and it constrains the interaction parameter $r$ almost symmetrically around the zero value. Adding low redshift measurements, $\lambda$ tends to its lower limit, while the interaction parameter breaks the symmetry around the zero value. We see that, including BAO, SNIa and $H_0$, the null interaction gets disfavored and the interaction parameter shows a preference for the negative value.

\begin{figure*}[htbp!]
\centering
    \includegraphics[scale=0.45,angle=-90]{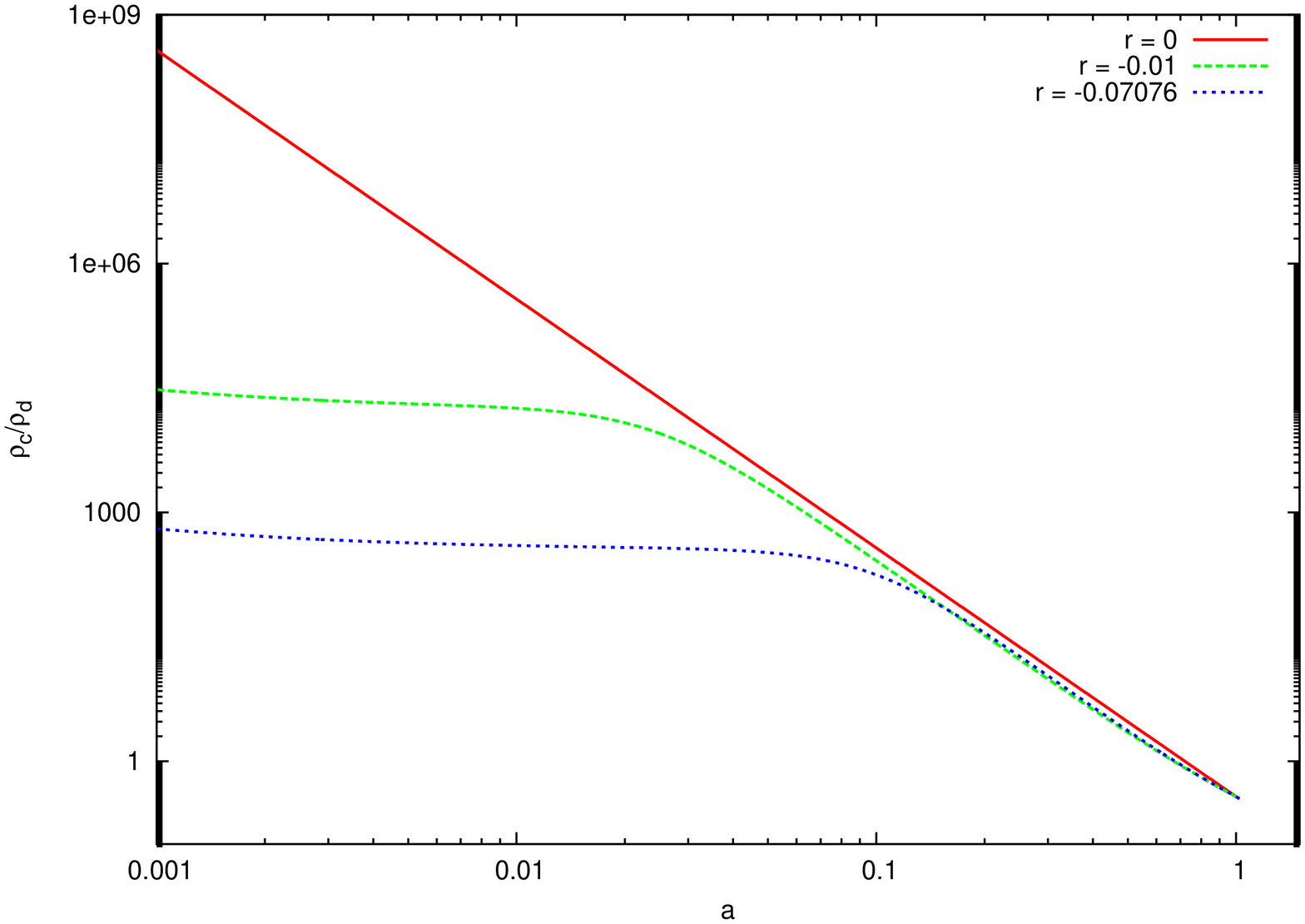}
\caption{(Color online). Evolution of the ratio between the energy densities of dark matter and dark energy.}\label{Qcouple_ratio}
\end{figure*}

We then consider the case when we fix the scalar potential parameter $\lambda$. We have learned that as we increase the value for $\lambda$, the interaction becomes more favored. For instance, $\lambda = \sqrt{3/2}$ produces the results in Table \ref{bestfit_lagrangian2}. The 1-D posterior distributions are plotted in Fig. \ref{1d_dist_lagrangian2}. These results show that even when we fix the parameter $\lambda$, Planck data are compatible with a null interaction. However, if we include low redshift measurements from BAO, SNIa and $H_0$, the data favor a negative value of $r$. For this value of $\lambda$, the negative interaction parameter is favored by more than 99\% C.L. when we consider all data sets. Augmenting the value of $\lambda$, a negative $r$ is even more favorable. Thus, we conclude that if we are able to determine the value of $\lambda$, or if we have a theoretical model fixing it, considering that this value is sufficiently large, the Yukawa interaction between dark energy and dark matter will be preferred by the cosmological data. Besides, the best fit value that we have obtained for the interacting parameter $r$ helps alleviating the coincidence problem because there is more time to the energy densities of dark matter and dark energy to be comparable. Figure \ref{Qcouple_ratio} shows the effect of the interaction in the ratio of the energy densities of dark matter and dark energy.

\section{Conclusions}
\label{sec:conclusions}

In this paper we have obtained cosmological
constraints on the Yukawa-type dark
matter-dark energy interaction model from the
new CMB measurements provided by the Planck
experiment. We have found that a dark coupling
interaction is compatible with Planck data, although they are still consistent with a null interaction.

We have also considered the combined constraints
from the Planck data plus other observations from low redshift measurements.
These extra data have induced evidence for a negative value of the interaction parameter $r$. When we allowed the scalar potential parameter $\lambda$ to vary freely, the interaction disfavor the null interaction, but still remained consistent with it. However, fixing $\lambda$, we obtained significantly evidence for interaction. For $\lambda = \sqrt{3}/2$, for example, we found a negative interaction at more than 99\% C.L., and higher values of $\lambda$ favor even more the interaction. Thus, we conclude that the Yukawa coupled dark energy
model is viable and is favorable for sufficiently high values of the scalar potential parameter. Besides, the best fit value that we obtained helps alleviating the coincidence problem.

\begin{acknowledgments}

A.C. and E.A. acknowledge financial support from
CNPq (Conselho Nacional de Desenvolvimento
Cient\'\i fico e Tecnol\' ogico), and E.A. and
L.O. also acknowledge FAPESP (Funda\c c\~ao de Amparo
\`a Pesquisa do Estado de S\~ao Paulo).

This work has made use of the computing facilities of the Laboratory of Astroinformatics (IAG/USP, NAT/Unicsul), whose purchase was made possible by the Brazilian agency FAPESP (grant 2009/54006-4) and the INCT-A.

\end{acknowledgments}

\bibliographystyle{apsrev4-1}
\bibliography{references}

\end{document}